# Geometrical control of interface patterning underlies active matter invasion

Haoran Xu [1], Mehrana R. Nejad [2], Julia M. Yeomans [2], Yilin Wu [1]

[1] *Department of Physics and Shenzhen Research Institute, The Chinese University of Hong Kong, Shatin, NT, Hong Kong, P.R. China.*
[2] *The Rudolf Peierls Centre for Theoretical Physics, Department of Physics, University of Oxford, Parks Road, Oxford OX1 3PU, United Kingdom*

**Abstract**

Interaction between active materials and the boundaries of geometrical confinement is key to many emergent phenomena in active systems. For living active matter consisting of animal cells or motile bacteria, the confinement boundary is often a deformable interface, and it has been unclear how activity-induced interface dynamics might lead to morphogenesis and pattern formation. Here we studied the evolution of bacterial active matter confined by a deformable boundary. We discovered that an ordered morphological pattern emerged at the interface characterized by periodically-spaced interfacial protrusions; behind the interfacial protrusions, bacterial swimmers self-organized into multicellular clusters displaying +1/2 nematic defects. Subsequently, a hierarchical sequence of transitions from interfacial protrusions to creeping branches allowed the bacterial active drop to rapidly invade surrounding space with a striking self-similar branch pattern. We found that this interface patterning is controlled by the local curvature of the interface, a phenomenon we denote as collective curvature sensing. Using a continuum active model, we revealed that the collective curvature sensing arises from enhanced active stresses near high-curvature regions, with the active length-scale setting the characteristic distance between the interfacial protrusions. Our findings reveal a protrusion-to-branch transition as a novel mode of active matter invasion and suggest a new strategy to engineer pattern formation of active materials.

**Introduction**

As a prototype of active matter [1], active fluids consist of force-generating units suspended in a liquid medium [2], such as suspensions of motile bacteria [3-5], animal tissues [6, 7], and cytoskeleton components driven by molecular motors [8, 9]. Interaction between active fluids and the boundaries of geometrical confinement has been shown to support rich emergent phenomena, such as cell sorting [10, 11], spontaneous flow [9, 12], vortex formation [13, 14] and ordering [15]. These emergent phenomena are primarily observed in active fluids interacting with hard confining walls. However, for living active fluids in natural environments or in clinical settings, the confinement boundary is often a soft and deformable interface. For example, the collective invasion of mesenchymal cancer cells pushes against soft tissue layers [16, 17]; surface-associated bacterial communities (or biofilms) can grow as droplets on solid substrates where the interfacial mechanics is key to their development [18, 19]. Activity-induced boundary deformation or interface growth enables active matter invasion into the surrounding space [20, 21] and it is a key process that governs the morphogenesis and pattern formation of active materials [22-25].

Motile bacteria are premier model systems for active matter [1, 2]. Here we studied the evolution of bacterial active fluids confined by a deformable boundary, i.e., the three-phase (air-liquid-solid) interface. Our experimental setup consists of drops of suspended motile bacteria grown on solid substrates; this configuration resembles the initial stages of growing bacterial colonies and biofilms commonly found in the natural environment and during pathogenesis [18]. We discovered that an ordered morphological pattern emerged at the interface, characterized by periodically-spaced interfacial protrusions. Behind the interfacial protrusions, bacterial swimmers self-organized into multicellular clusters each displaying a +1/2 nematic defect, and the collective velocity field of cells adjacent to the invasion interface was organized into a 1D vortex lattice. Some of the interfacial protrusions eventually broke out as creeping branches, and the branches continued to experience the protrusion-to-branch transition in a hierarchical manner, allowing the bacterial active fluid to rapidly invade the surrounding space with a striking self-similar branch pattern. We found that the interface patterning of active suspension drops is controlled by a mechanism we denote as collective curvature sensing: both the interfacial protrusion amplitude and the probability of the protrusion-to-branch transition are positively correlated with the local curvature of the drop boundary.  We developed a continuum active matter model that reproduces the interface morphology and revealed that the collective curvature sensing arises from enhanced active forces near high-curvature regions.   Our findings illustrate the key role of interface geometry in active matter invasion and pattern formation.  The protrusion-to-branch transition as a novel mode of space exploration is also relevant to the expansion and dispersal of bacterial communities in natural environments [18].

**Results**

**Ordered protrusions and self-organization at an active fluid interface**

We used quasi-2D circular drops containing a suspension of *Proteus mirabilis* grown on an agar substrate as a model active fluid (Methods).  *P. mirabilis* is a Gram-negative, rod-shaped bacterium, widely distributed in the natural environment [26].  It is well known for having vigorous flagellar motility on agar surfaces [27].  We monitored how the active suspension drops of *P. mirabilis* started to spread on solid substrates with low wettability (Methods).  Initially, the interface between the active drop and the cell-free space was smooth. As time evolved, the interface was gradually deformed, and the active drop developed an ordered morphological pattern at the interface (Fig. 1a; Fig. 1b, upper; Video S1).  This was characterized by periodically-spaced protrusions along the interface, with an inter-protrusion separation of 65±9 μm (mean±S.D.)  (Fig.1c).  Behind the interfacial protrusions, bacterial swimmers self-organised into multicellular clusters, commensurate with the protrusions, and with a length (from base to tip) of ~10-30 μm.  The orientation field of cells in each cluster displayed a +1/2 nematic defect with the polarity vector [28] pointing perpendicular to the interface and towards the colony center (Fig. 1b, middle; Video S1).  Meanwhile, the collective velocity field of cells adjacent to the interface self-organized into a vortex lattice, with neighboring domains of opposite vorticity (Fig. 1b, lower).  The vortices were primarily extended perpendicular to the interface, and the domain size was, on average, 1/2 of the inter-protrusion separation (Fig. 1b, lower).  The vortex lattice was highly stable in time, as shown by the strong temporal correlations (Fig. 1d).

To examine whether the interface morphological pattern and bacterial self-organization was driven by bacterial motility, we optically deactivated the flagellar motility of cells at the invasion interface with violet light illumination [14] (Methods). During light-induced motility deactivation, the interfacial protrusions gradually smoothed out and cells near the edge became aligned parallel to the interface (Fig. S1; Video S2). Meanwhile, all the multicellular clusters as well as the vortex structure of the cellular collective velocity field near the interface disappeared. After withdrawal of light, cell motility recovered and the interfacial protrusions re-emerged (Fig. S1; Video S2). As the growth activity of most cells in the active drop were not affected during motility deactivation (note that only cells at the edge were illuminated by light), our result showed that the interface morphological pattern and cellular self-organization do not depend on growth pressure [29], but rather are driven by active stress due to flagellar motility. On the other hand, bacterial suspension drops consisting of cells that were washed and deposited on fresh substrates developed the same interfacial morphology within ~10 min (Fig. S2; Video S3; Methods); this fact rules out the contribution of Marangoni instability [30-33], chemotaxis [34], or intercellular chemical signaling [35] to the phenomena. Finally, we note that the protrusion-to-branch pattern of colony dispersal is distinct from the fingering pattern of bacterial colonies driven by Marangoni instability in the presence of surface tension gradients [31-33], which requires biosurfactant synthesis and is independent of cell motility.

**Hierarchical transition from interfacial protrusions to creeping branches**

After the emergence of the interfacial pattern, some protrusions of the active suspension drop started to expand continuously at a speed of several µm/s (up to ~10 µm/s) in the form of creeping branches (Fig. 2a; Video S4). The first creeping branches later underwent the same protrusion-to-branch transition process (Fig. 2a): Regularly spaced protrusions developed at the interface between the creeping branches and the open space, and some of these protrusions broke out as secondary creeping branches with a narrower width. The secondary creeping branches again repeated this process to give rise to new creeping branches at the next level. The protrusion-to-branch transition continued recursively until the newly formed branches were too narrow to host the cellular clusters that are associated with interfacial protrusions (Video S5). We note that the width of creeping branches could increase due to more cells replenishing them from the main suspension drop or due to the growth of cells. Also, two creeping branches colliding with each other would merge into one. Not all protrusions eventually develop into creeping branches; as a protrusion started to expand, we found that the amplitude of nearby protrusions would decrease and even disappear (Fig. S3), suggesting that cells are being drawn to support the expansion of a nearby existing creeping branch.

The recursive and hierarchical transition process gives rise to a striking self-similar pattern (Fig. 2b; Video S5). To characterize this pattern, we computed its fractal dimension $d$, ~1.70 (Fig. 2c) (Methods) [36]. The fractal dimension is close to that reported for branching patterns of bacterial colonies [37, 38]. However, here the self-similar branching pattern of the active suspension drop is driven by motility-induced active stresses, which is in contrast to the fractal branching patterns of bacterial colonies due to reaction-diffusion processes involving the cells and nutrient or chemoattractant fields [37, 38].

**Interface curvature controls protrusion amplitude and the probability of a protrusion-to-branch transition**

When working with active suspension drops of an anisotropic shape, we noticed that the first creeping branches were more likely to emanate from higher-curvature regions (Fig. 3a). To rationalize this observation, we investigated the behavior of interfacial protrusions in active suspension drops with different shapes. We discovered that the amplitude of the interfacial protrusions was positively correlated with the baseline curvature of the interface (i.e., the local curvature computed by smoothing out interfacial protrusions) (Fig. 3b). For instance, the protrusion amplitude near the poles (of higher curvature) of an elongated drop was significantly greater than elsewhere while the inter-protrusion separation remained the same (Fig. 3c). Moreover, examination of the full hierarchical branching pattern of active suspension drops revealed that the probability of a protrusion-to-branch transition increased with the baseline curvature of the interface (Fig. 3d; Methods). These results show that the dynamics of protrusions is controlled by the curvature of the interface of active suspension drops and branches, a phenomenon we denote as collective curvature sensing.

The light-induced motility deactivation experiment described above (Fig. S1) suggests that the protrusion amplitude is related to the magnitude of active stress generated by cells. Therefore, the collective curvature sensing may result from active stress heterogeneity in the active suspension drop. To examine this idea, we measured the collective cellular speed behind the boundary (ranging from ~20 to ~50 µm from the interface) at different locations within anisotropic active drops; collective speed increases with active force in quasi-2D active bacterial suspensions in contact with a frictional substrate. Indeed, we found that the bacteria had a higher collective speed behind the boundary regions with greater curvature while the cell density was almost homogeneous (Fig. 3e; Fig. S4). The curvature-dependence of the active-stress distribution also explains the differential probability of protrusion-to-branch transitions: At boundary regions with a higher baseline curvature, the higher collective active stress can more easily overcome the liquid-air interfacial tension, thus giving rise to higher probabilities of branch invasion.

**Numerical simulations explain the interface morphology and collective curvature sensing**

Our experiments suggest that the interfacial patterning is likely driven by active stresses produced by motile bacteria in the suspension drop. To examine this idea, we ran continuum simulations describing the suspension droplet as a 2D active nematic fluid and measuring the collective velocity and nematic order parameter (Supplemental Information). Guided by the experiments the strength of nematic order was taken to decay from the edge to the center. As initial conditions we took the director field parallel to the interface with an imposed noise and zero velocity field, mimicking the parallel anchoring of cells at the interface in experiments. Extensile active fluids are unstable, and at early times a bend instability formed around the drop, thus initiating protrusions at approximately regular positions around the interface (Fig. 4A; Fig. 4B, top). The director configuration within the protrusions led to outwards radial flow which in turn increased the definition of the protrusions and resulted in a director field resembling a +1/2 topological defect within each protrusion, as observed in the experiments (compare Fig. 4B, top and Fig. 1B, middle). Note also the alternating planar (between the protrusions) and perpendicular (under the protrusions) director field at the edge of the main drop. We then mimicked the very low velocity of the bacteria within the protrusions by reducing the activity to zero within the protrusions. The alternating director field at the drop edge acted as boundary conditions to stabilize a flow-vortex lattice with the same structure as in the experiments (compare Figs. 4B, bottom and 1B, bottom). This was a persistent state until we re-introduced a small activity within the protrusions. They then slowly grew and eventually active material escaped from certain of the protrusions, as the creeping branches did in the experiments.

The agreement between the phenomena in experiments and continuum modeling supports the notion that the bacterial suspension drop can be described as an active drop whose dynamics is driven by cell motility. Our simulations suggest that the regularity of protrusion distribution along the interface is rooted in the bend instability. Indeed, the simulations showed that the distance between the protrusions does not depend on system size or curvature but is set by the active length-scale $\sqrt{K_Q/\zeta}$, where $K_Q$ is the Frank elastic constant and $\zeta$ is an activity constant (Fig. 4c; Supplemental Information). Further, to explain collective curvature sensing as found in the experiment, we ran simulations of an elliptical active drop. In agreement with the experiment, the result shows that the protrusions first start growing in the regions of higher curvature (Fig. 4D), with the velocity near the interface of the elliptical drop proportional to curvature (Fig. 4E). This result can be understood as follows in the framework of our continuum model. The active force is proportional to the divergence of the Q-tensor, and is increased by the bend distortion imposed on the director field by a curved interface. This is due to the parallel alignment of the director with the interface, promoted by both the anchoring term in the free energy and the extensile stress [39]. Since the director is aligned with a curved surface, it deviated from the nematic alignment and forms a bend that produces flows. Moreover, we looked at the dynamics of the interface at large times, and found that, similarly to the experiment, when branches become elongated enough, they undergo another bend instability and form new branches (Fig. 4F-G; Video S6). Although numerical resolution prevents us from reproducing the large number of consecutive protrusion-to-branch transitions observed experimentally, this result provides additional evidence that the fractal pattern is created by consecutive bend instabilities due to active flows.

**Discussion**

Taken together, we have discovered a novel mode of active matter invasion mediated by the interaction between activity-induced flows and deformable interfaces. In this mode, active drops of bacterial suspensions develop regularly-spaced interfacial protrusions at the edge; the protrusions further transform into creeping branches in a hierarchical, self-similar manner, with the transition probability depending on the local interface curvature. Continuum simulations of an active drop model reveal that the highly ordered pattern of interfacial protrusion is initiated by bend instabilities, and that enhanced active forces near high-curvature regions result in curvature-dependent protrusion dynamics which we term collective curvature sensing.

Self-organized pattern formation is a hallmark of living systems and a promising route to engineer functionalities of living materials [40]. Ranging from bacterial colony development [38, 41] to animal embryogenesis [42], the formation of biological spatial patterns generally requires a complex interplay of genetic regulation, inter-cellular communication, and mechanical feedback [43-45]. By contrast, the ordered protrusion and branching pattern reported here relies on purely physical interactions between activity-induced flows and deformable interfaces. In particular, the phenomenon of collective curvature sensing shows that interface geometry may play important roles in active matter morphogenesis. It suggests a novel strategy, i.e., designing curvature cues at the interface of active fluids, to guide and manipulate pattern formation in active materials.

While we expect that the mode of ordered invasion and pattern formation applies to nematic active fluids in general, our findings are of direct relevance to the expansion and dispersal of bacterial communities. Bacterial communities in ecological and clinical settings are commonly found in interface-associated environments, such as biofilms [46, 47] and bacterial swarms [18, 48] that develop on solid substrates. If the substrates are highly wettable, communities of motile bacteria can establish a thin liquid film and explore new space rapidly by a process known as swarming, in which flagellar motility drives uniform expansion of the liquid film [49, 50]. However, swarming is inhibited on substrates with low wettability (e.g., high-concentration agar substrates in laboratory settings or substrates without wetting agents). Under such circumstances, a protrusion-to-branch transition becomes a viable solution for bacterial communities to explore new territories.

**Methods**

No statistical methods were used to predetermine sample size.

**Bacterial Strains.** The following two strains were used: wildtype *P. mirabilis* BB2000, and a fluorescent *P. mirabilis* KAG108 [BB2000 background with constitutive expression of Green Fluorescent Protein (GFP) and with an ampicillin resistance marker [51]; from Karine Gibbs, Harvard University, Cambridge, MA]. Single-colony isolates were grown overnight (~13–14 h) with shaking in LB medium (1% Bacto tryptone, 0.5% yeast extract, 0.5% NaCl) at 30 °C to stationary phase. For *P. mirabilis* KAG108, ampicillin (100 μg/mL) was added to the growth medium to maintain the plasmid.

**Agar plates.** LB agar (Difco Bacto agar at specified concentrations infused with 1% Bacto tryptone, 0.5% yeast extract, 0.5% NaCl) was autoclaved and stored at room temperature. Before use, the agar was melted in a microwave oven, cooled to ~60 °C, and pipetted in 10-mL aliquots into 90-mm polystyrene Petri plates. The plates were swirled gently to ensure surface flatness, cooled for 10 min without a lid inside a large Plexiglas box, and then covered by the lid for further experimentation. Two approaches, namely colony growth and direct deposition, were adopted to prepare bacterial active suspension drops containing *P. mirabilis*.

**Microscopy imaging.** All imaging was performed on a motorized inverted microscope (Nikon TI-E). The phase-contrast images of bacterial active suspension drops, interfacial protrusions, and creeping branches were acquired with 20× (Nikon S Plan Fluor 20×, numerical aperture 0.45, working distance 8.2-6.9 mm), 10× (Nikon CFI Achromat 10×, numerical aperture 0.25, working distance 7.0 mm) or 4× (Nikon Plan Fluor 4×, numerical aperture 0.13, working distance 16.5 mm) objectives; the images were recorded by a scientific complementary metal-oxide-semiconductor (sCMOS) camera (Andor Zyla 4.2 PLUS USB 3.0) at 10 fps and at full frame size (2048×2048 pixels). To tune the speed of cells in the bacterial suspension, cells were illuminated by violet light of ~1740 mW/cm$^2$ provided by Nikon Intensilight and passing through the 20× objective via a 406 nm filter (406/15 nm; FF01-406/15-25, Semrock Inc.). To assess bacterial number density in active suspension drops, GFP-tagged *P. mirabilis* KAG108 cells were imaged in epifluorescence using the 20× objective and an FITC filter cube

(excitation 482/35 nm, emission 536/40 nm, dichroic 506 nm; Semrock Inc.), with the excitation light provided by a mercury precentred fibre illuminator (Nikon Intensilight); the fluorescence images were recorded with the sCMOS camera (Andor Zyla 4.2 PLUS USB 3.0). In all experiments, the Petri dishes were covered with a lid to prevent evaporation and air convection. The sample temperature was maintained and controlled via a custom-built temperature-control system installed on the microscope stage.

**Image processing and data analysis.** Images were processed using the open-source Fiji (ImageJ) software (http://fiji.sc/Fiji) and custom-written programs in MATLAB (The MathWorks, Inc.). Prior to processing, the microscopy images were smoothed to reduce noise by convolution with a Gaussian kernel of standard deviation ~1 μm. To acquire the bacterial orientation field, we first calculate the structure tensor defined as

$$J(x,y) = \begin{bmatrix} I_x I_x & I_x I_y \\ I_y I_x & I_y I_y \end{bmatrix},$$

where $I_x = \partial_x I$, $I_y = \partial_y I$ and $I = I(x,y)$ is the light intensity field of a phase-contrast image [52]. As the intensity of the regions occupied by bacterial cells is different than the background in phase-contrast imaging, the spatial gradients of the intensity field that appear in the structure tensor contains information of the spatial distribution of cell orientations: for a given position $(x, y)$, the orientation $\theta(x, y)$ of the eigenvector of the structure tensor $J(x, y)$ with greater eigenvalue typically corresponds to the local orientation of cells located at the position. The bacterial orientation field was obtained by coarse-graining the local orientations $\theta(x, y)$ with a grid size of 10 pixel × 10 pixel (3.25 μm × 3.25 μm); the mean orientation at a grid, denoted as $\langle \theta \rangle$, is computed as $\arg[\sum_k \exp(i\theta_k)]$ [53], where arg(x) denotes the argument of the complex number x. To identify the nematic defects near the interface, we computed the net change of director angle along a counterclockwise loop around each grid, and grids with ~π director angle change were identified as the cores of +1/2 nematic defects [54]. The direction of a +1/2 nematic defect is calculated as $\boldsymbol{n}_i = \nabla \cdot J(\boldsymbol{r}_i)/|\nabla \cdot J(\boldsymbol{r}_i)|$, where $r_i$ is the location of the defect core [28].

To identify the protrusions at the interface of the active suspension drop, we first converted the phase-contrast image to a binary image to identify the interface. For circular active suspension drops, the active fluid interface was represented as $\rho = f(\phi)$ in polar coordinate with the origin set as the center of the active suspension drop. For active suspension drops with an anisotropic shape, the interface was divided into ~200-μm segments; each segment was represented as $\rho = f(\phi)$ in a local polar coordinate system whose origin was chosen as the center of curvature of a circular arc that best fitted the interface segment. Next, the active fluid interface (or interface segment) was smoothed along $\phi$ to reduce noise by convoluting $f(\phi)$ with a Gaussian kernel of standard deviation of $\Delta\phi$=1 μm/$\bar{\rho}$. The protrusions and valleys were identified as $\partial_\phi \rho = 0$ and $\partial_\phi^2 \rho < 0$ (for protrusions) or $\partial_\phi^2 \rho > 0$ (for valleys). The amplitude of the interfacial protrusions was defined as half of the difference in averaged polar radius between protrusions and valleys, i.e., $(\bar{\rho}_{protrusion} - \bar{\rho}_{valley})/2$. To calculate the baseline curvature of the active fluid interface or creeping branches, segments of interface or creeping branches with ~200 μm in length were fitted to circular arcs via the least squares method in MATLAB.

To compute the collective velocity near the interface, we first performed optical flow analysis based on phase-contrast time-lapse videos using the built-in OpticalFlowHS function [55] of MATLAB with a grid size of 1 pixel × 1 pixel. Prior to the optical flow analysis, the microscopy images were smoothed to reduce noise by convolution with a Gaussian kernel of standard deviation of 1 μm. The results were insensitive to different parameters of smoothing. The optical flow analysis yielded a space- and time-dependent collective velocity field $\vec{v}(\vec{r}, t)$, which was then decomposed in polar coordinates as $(v_r(\vec{r}, t), v_\phi(\vec{r}, t))$ ($v_r$ denotes radial component and $v_\phi$ denotes tangential or azimuthal component; the origin of the polar coordinate system is chosen at the center of the circular bacterial active suspension drop). The spatial-temporal autocorrelation of radial velocity near the interface is computed as $C(\Delta\phi, \Delta t) = \frac{<v_r(R,\phi,t)v_r(R,\phi+\Delta\phi,t+\Delta t)>_{\phi,t}}{<v_r(R,\phi,t)^2>_{\phi,t}}$, where angular brackets $\langle ... \rangle_{\phi,t}$ indicate averaging over the azimuthal angle and time $t$, and $R$ is the averaged radius of the active suspension drop.

To calculate the fractal dimension $D_f$ of creeping branch patterns, we adopted the box-counting method [36-38]. Specifically, an image of creeping branches was first divided into a lattice of square domains of size $L \times L$. The number of squares that covered the pattern of interest (i.e., creeping branches) were counted and denoted as $N(L)$. This procedure was repeated for different choices of $L$, and the fractal dimension $D_f$ of the pattern was obtained with the overlapping formula $N \sim L^{-D_f}$.

**Supplementary Materials**, including Supplemental Information with details of the theory and Supplementary Videos, is available in the online version of the paper.

**Data availability.**  The data supporting the findings of this study are included within the paper and its Supplementary Materials.

**Code availability.**  The custom codes used in this study are available from the corresponding author upon request.

**Acknowledgements**.  We thank Karine Gibbs (Harvard University) for providing the bacterial strains.  This work was supported by the National Natural Science Foundation of China (NSFC No. 31971182, to Y.W.), the Research Grants Council of Hong Kong SAR (RGC Ref. No. RFS2021-4S04, 14306820, 14306820; to Y.W.).  M.R.N. acknowledges the support of the Clarendon Fund Scholarships.

Figure 1

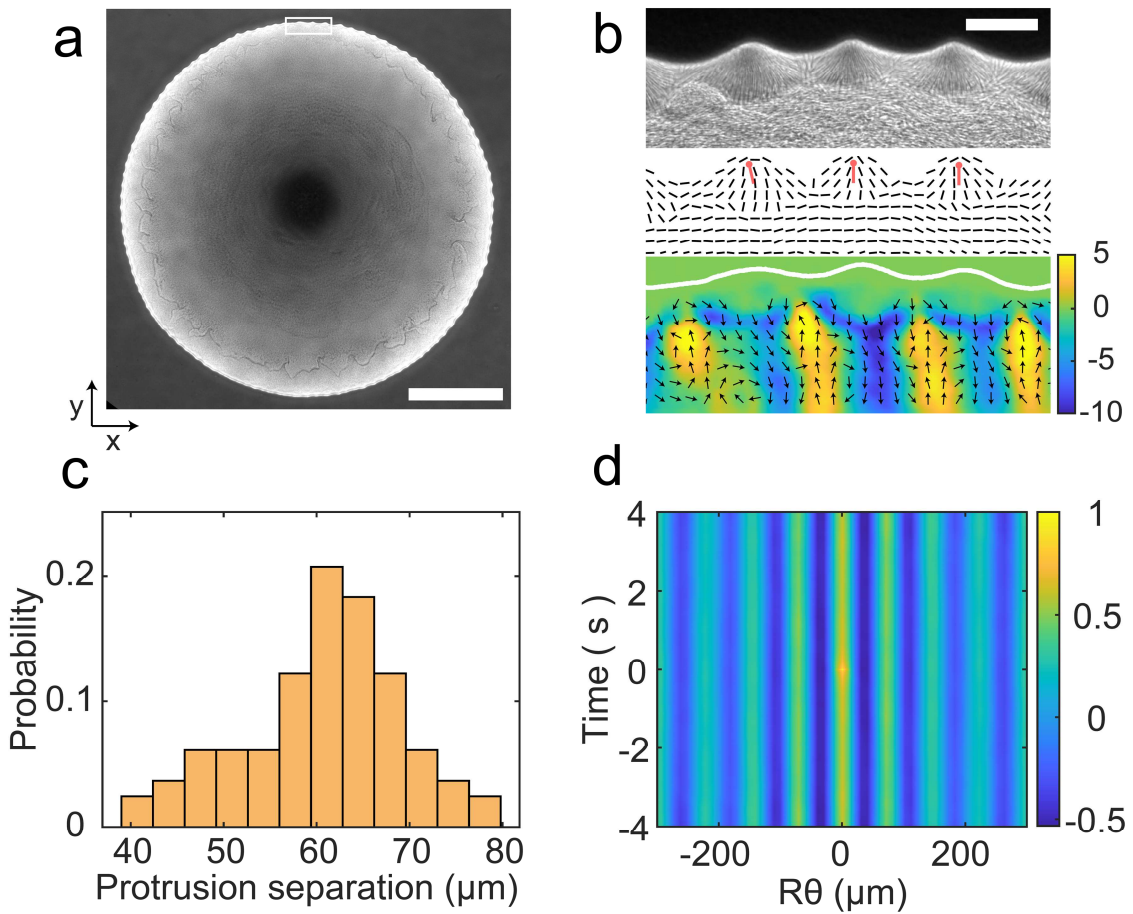

**Fig. 1. Ordered interfacial protrusions and bacterial self-organization near an active fluid interface.** (**a**) Representative phase-contrast image of a *P. mirabilis* active suspension drop with ordered interfacial protrusions. Scale bar, 500 μm. (**b**) Upper: enlarged view of the white box in panel a. Middle: orientation field of cells in the upper panel computed based on the gradient field of the phase-contrast image (Methods). Red marks represent the direction of +1/2 defects, with the dot indicating the defect core. Lower: Colormap showing the averaged velocity field at the region shown in the upper panel computed by optical flow analysis on phase-contrast images (Methods). The collective velocity field was averaged over a duration of 10 s. Arrows in the figure

represent velocity direction while the colormap at the right indicates the radial velocity component (unit: µm/s). Positive value indicates moving outwards toward the interface while a negative value indicates moving inwards away from the interface. Scale bar, 50 µm. (**c**) Distribution of the nearest separation of interfacial protrusions. (**d**) Spatial-temporal autocorrelation of the radial velocity adjacent to the interface. On the horizontal axis R represents the radius of the active suspension drop and θ is the polar angle.

Figure 2

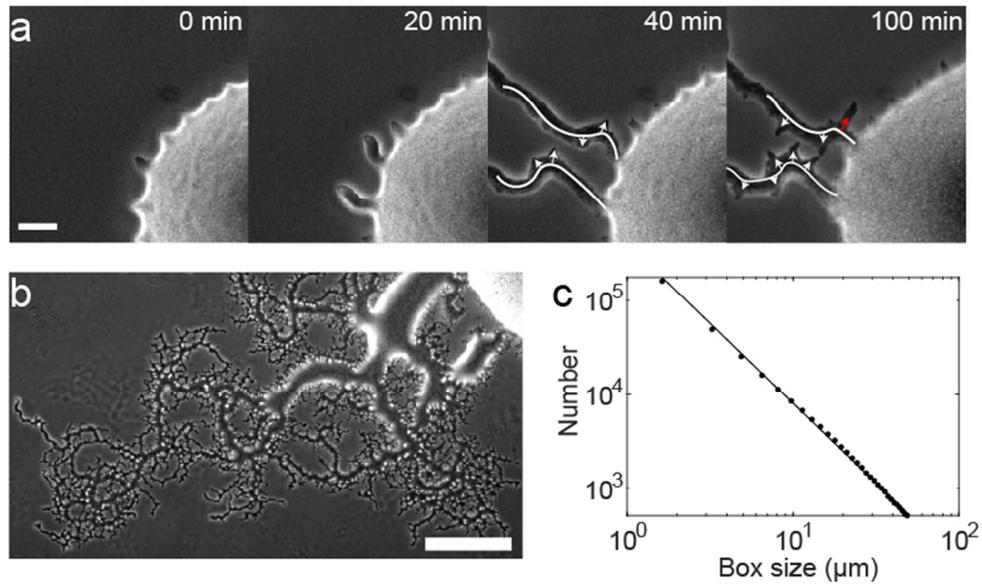

**Fig. 2 Hierarchical transition from interfacial protrusions to creeping branches.** (**a**) Image sequence showing the transition from interfacial protrusions to creeping branches. Scale bar, 50 μm. T= 0 min is chosen at the onset of the protrusion-to-branch transition. White arrows on the T= 40 min and 100 min images represent the protrusions formed on the first branches, and the red arrow at T=100 min indicates a position where a secondary creeping branch is forming. (**b**) Self-similar pattern resulting from the recursive and hierarchical protrusion-to-branch transition process. The brighter region at the upper-right corner is the edge of the active suspension drop. Scale bar, 500 μm. (**c**) Fractal analysis of the branch pattern by the box counting method (Methods). Data in the log-log plot is fitted by a straight line, suggesting that the branch pattern has self-similarity. The absolute value of the slope is taken as the fractal dimension (Methods).

Figure 3

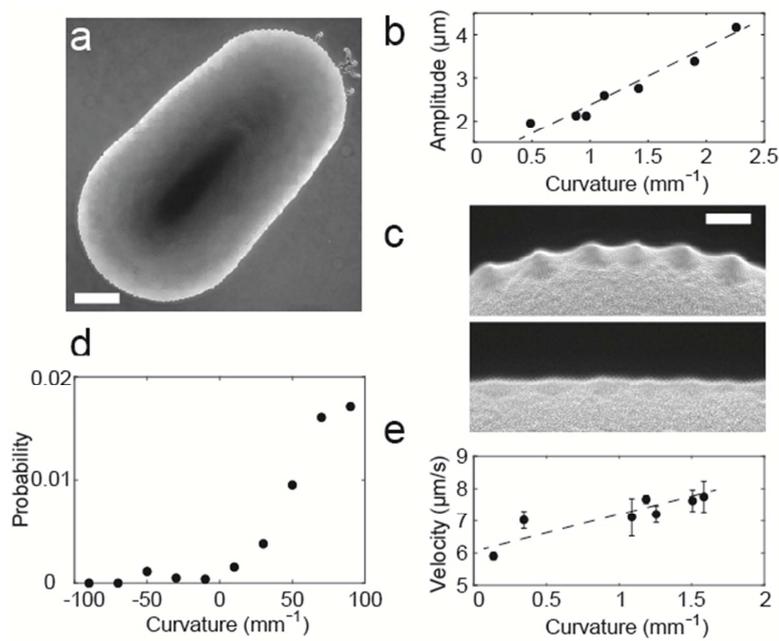

**Fig. 3. Curvature dependence of protrusion amplitude and protrusion-to-branch transition probability.** (**a**) Phase-contrast image of an elongated *P. mirabilis* suspension drop with creeping branches appearing only at the pole at the upper right corner, where the baseline curvature is relatively high. Scale bar, 500 μm. (**b**) Amplitude of interfacial protrusions plotted against the baseline curvature of the active suspension drop (Methods). (**c**) Enlarged view of the pole (upper) and the lateral side (lower) regions in panel a. Scale bar 50 μm. (**d**) Probability density function of a protrusion-to-branch transition at the interface of active suspension drops or creeping branches plotted against baseline curvature (Methods). Note that the curvature magnitude of creeping branches can be much larger than that of active suspension drops. (**e**) Collective speed of cells behind the interface of active suspension drops (ranging from ~20 to ~50 μm from the boundary) plotted against baseline curvature.

Figure 4

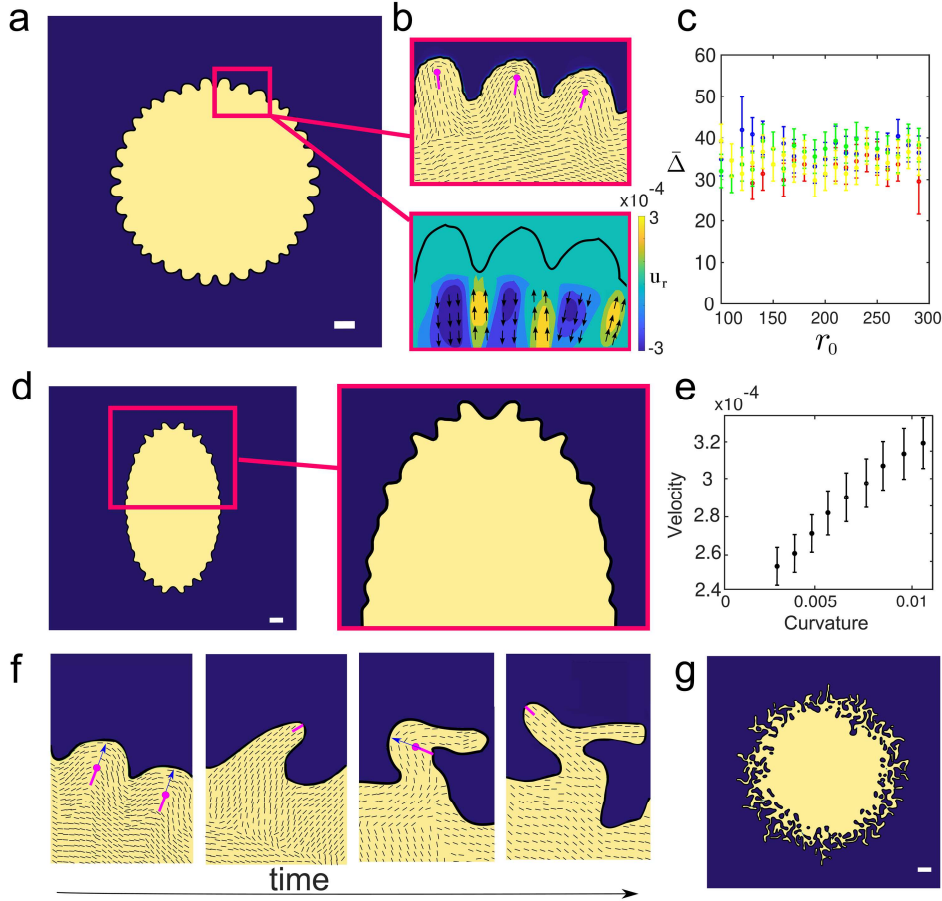

**Fig. 4. Deformation of an active droplet due to extensile active stress.** All the scale bars are 15 lattice Boltzmann units. (**a**) A periodic pattern of protrusions forms at the interface of the droplet. (**b**) Enlarged view of the pink box in panel a. Top: Orientation field around the protrusions. Magenta marks represent +1/2 defects. Bottom: Color map showing the averaged radial velocity field. Arrows indicate velocity direction. A positive value indicates moving outwards toward the interface while a negative value indicates moving inwards away from the interface. (**c**) The distance between the protrusions $\bar{\Delta}$ (rescaled with active length-scale $\sqrt{K_Q/\zeta}$) as a function of drop radius $r_0$. Color of data points indicates different active length-scales (blue: $K_Q = 0.005, \xi = 0.02$; green: $K_Q = 0.01, \xi = 0.02$; yellow: $K_Q = 0.01, \xi = 0.01$; red: $K_Q = 0.005, \xi = 0.01$). The collapse of data in different colors shows that $\bar{\Delta}$ is set by the active length-scale. (**d**) Formation of protrusions in an elliptical droplet. The protrusions first start growing in regions with higher curvature (poles). (**e**) Average velocity at the interface of the elliptical droplet as a function of the curvature, before the formation of defects. (**f**) As time passes, +1/2 defects move and form branches (elongated arms with nematic order). The elongated arms can then undergo a bend instability and form another arm. (**g**) At very large times, all the arms have undergone bend instabilities, forming a pattern with many branches.

**Supplementary Figures**

Figure S1

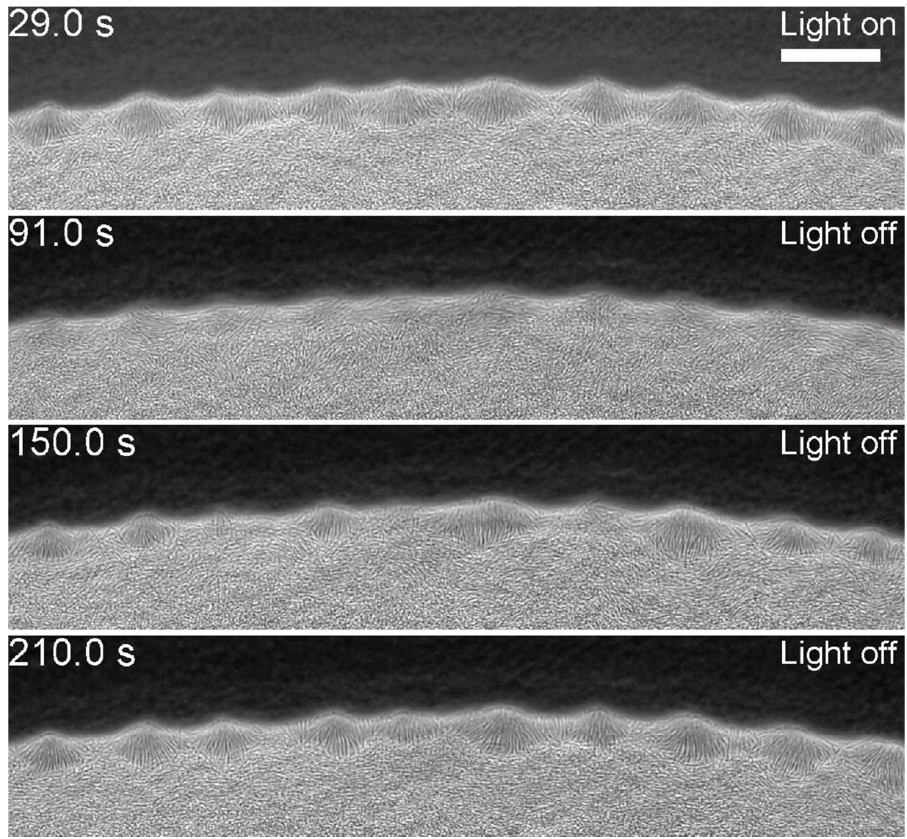

**Fig. S1. Image sequence showing the effect of light-induced motility deactivation on interfacial protrusions.** A *P. mirabilis* active suspension drop was illuminated by violet light at T=29 s and the light was turned off at T=91 s. The motility of all cells in the entire field of view was suppressed by violet light illumination. The interfacial protrusions gradually smoothed out within ~20 s of violet light illumination (T≈49 s) and recovered within ~60 s after withdrawal of the violet light (T≈150 s). Scale bar, 50 μm. Also see Video S2 for the full process.

Figure S2

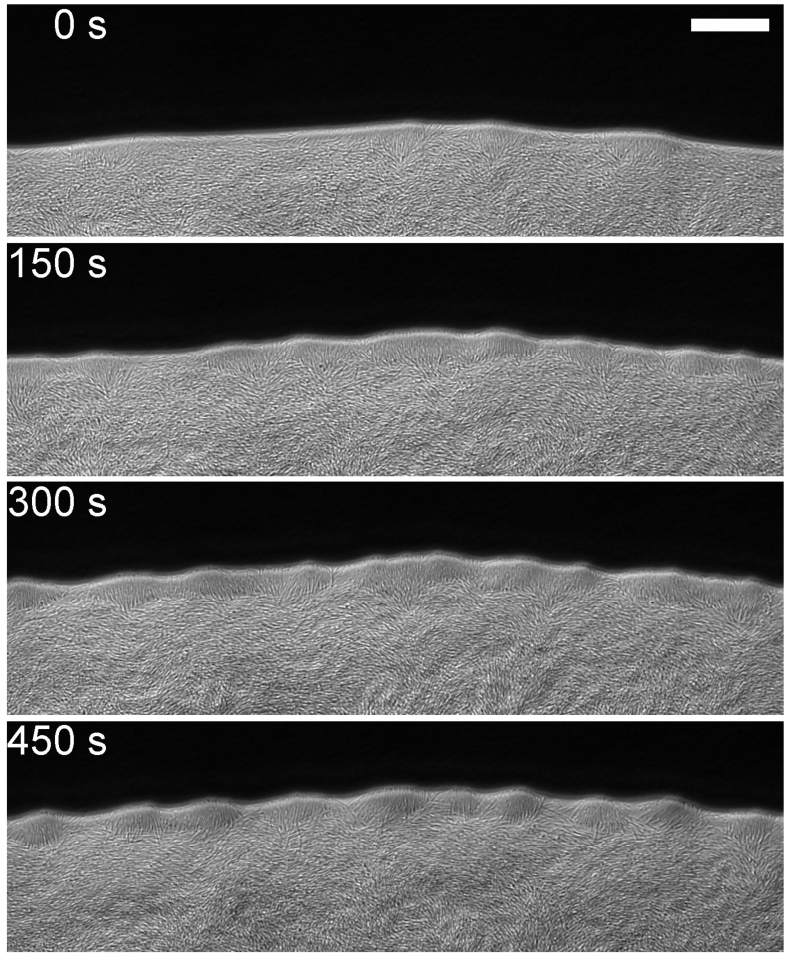

**Fig. S2. Image sequence showing the emergence of interfacial protrusions in *P. mirabilis* suspension drops prepared by direct deposition.** Following the "direct deposition" approach described in Methods, cells were collected from bacterial suspension drops, washed by centrifugation, and resuspended to a volume fraction of ~10%. Interfacial protrusions appeared within ~5 min of depositing the active suspension drops. Scale bars, 50 µm. Also see Video S3.

Figure S3

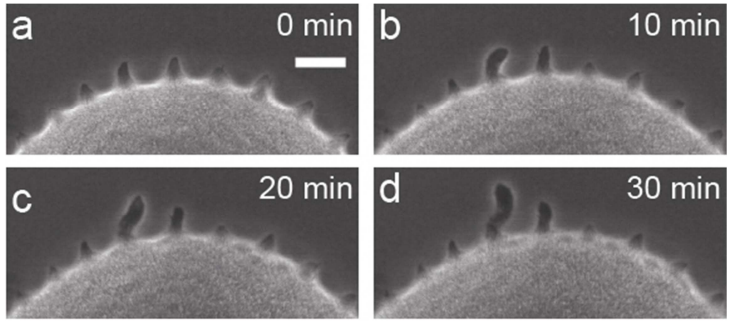

**Fig. S3. Interaction between creeping branches and nearby interfacial protrusions.** The image sequence shows that as protrusions in a *P. mirabilis* suspension drop started to expand, the amplitude of nearby protrusions would decrease and even disappear. Scale bar, 50 µm. The bacterial suspension drop was prepared by the colony-growth approach (Methods).

Figure S4

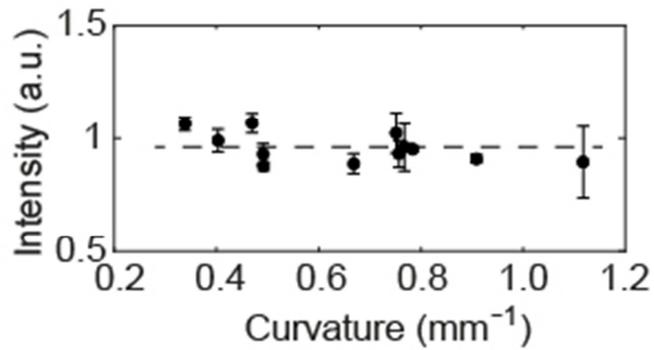

**Fig. S4. Curvature dependence of cell density near the interface of *P. mirabilis* suspension drops.** Fluorescence intensity of GFP-labeled cells in the suspension drop serves as a proxy of the surface-packing density of bacteria (Methods). The fluorescence intensity of cells was measured at ~20-50 μm radially inwards from the interface, and the curvature in the plot was the baseline curvature of the interface (i.e. local curvature computed by smoothing out interfacial protrusions). Error bars indicate standard deviation (N=3).

# Supplemental Information: Geometrical control of interface patterning underlies active matter invasion

## SIMULATION METHODS

### A. Model

We use the active nematohydrodynamic equations, the fundamental continuum equations that describe wet active nematics [1–4]. These are coupled equations for the evolution of the bacteria concentration field $\phi$, nematic tensor, $\mathbf{Q} = S(\mathbf{nn} - \mathbf{I}/2)$ in two dimensions, and the associated incompressible fluid velocity, $\mathbf{u}$. They read

$$\partial_t \mathbf{Q} + \mathbf{u} \cdot \boldsymbol{\nabla} \mathbf{Q} - \boldsymbol{\mathcal{W}} = \gamma \, \mathbf{H}, \tag{1}$$

$$\rho \left(\partial_t + \mathbf{u} \cdot \boldsymbol{\nabla}\right) \mathbf{u} = \boldsymbol{\nabla} \cdot \boldsymbol{\Pi} - \Gamma_f \, \mathbf{u}, \quad \boldsymbol{\nabla} \cdot \mathbf{u} = 0, \tag{2}$$

$$\partial_t \phi + \mathbf{u} \cdot \boldsymbol{\nabla} \phi = \Gamma_\phi \nabla^2 \mu, \tag{3}$$

$$\mu = \frac{\partial f}{\partial \phi} - \boldsymbol{\nabla} \cdot (\frac{\partial f}{\partial \boldsymbol{\nabla} \phi}), \tag{4}$$

$$f = \frac{C}{2}(\mathcal{J}\phi - Q_{ij}Q_{ij}/2)^2 + \frac{A}{2}\phi^2(1-\phi^2) + \frac{K_\phi}{2}\nabla_m\phi\nabla_m\phi + \frac{K_Q}{2}\nabla_m Q_{ij}\nabla_m Q_{ij} + a\nabla_i\phi \, Q_{ij} \, \nabla_j\phi. \tag{5}$$

In the definition of the nematic tensor the director field $\mathbf{n}$ represents the orientation of the nematic alignment and the magnitude of the nematic order is denoted by $S$. In the evolution of the $\mathbf{Q}$ tensor, $\gamma$ is the rotational diffusivity and the molecular field, $\mathbf{H} = -\partial f/\partial \mathbf{Q} + \boldsymbol{\nabla} \cdot (\partial f/\partial \boldsymbol{\nabla} \mathbf{Q})$, drives the nematic tensor towards the minimum of the free energy (5). The generalized advection term

$$\boldsymbol{\mathcal{W}} = (\lambda \mathbf{E} + \boldsymbol{\Omega}) \cdot (\mathbf{Q} + \frac{\mathbf{I}}{2}) + (\mathbf{Q} + \frac{\mathbf{I}}{2}) \cdot (\lambda \mathbf{E} - \boldsymbol{\Omega}) - \lambda(\mathbf{Q} + \mathbf{I})Tr(\mathbf{Q} \cdot \mathbf{E}) \tag{6}$$

models the response of the nematic field to the strain rate $\mathbf{E}$ and vorticity $\boldsymbol{\Omega}$, and $\lambda$ denotes the flow-aligning parameter. In the Navier-Stokes equation (2), $\rho$ is the density of the suspension, $\Gamma_f$ is a friction, and the stress tensor, $\boldsymbol{\Pi}$, includes viscous, elastic and active contributions. The viscous stress, $\boldsymbol{\Pi}^v = 2\eta \mathbf{E}$, where $\eta$ is the viscosity, and the elastic stress,

$$\Pi^p_{ij} = - P\delta_{ij} + \lambda(Q_{ij} + \delta_{ij})Q_{kl}H_{kl} - \lambda H_{ik}(Q_{kj} + \frac{\delta_{kj}}{2})$$
$$- \lambda(Q_{ik} + \frac{\delta_{ik}}{2})H_{kj} + Q_{ik}H_{kj} - H_{ik}Q_{kj} - K(\partial_i Q_{kl})(\partial_j Q_{kl}), \tag{7}$$

where $P$ is the pressure. These are familiar terms that appear in the dynamical equations of passive liquid crystals. Activity enters through an active stress $\boldsymbol{\Pi}^a = -\zeta \mathbf{Q}$ where, for the extensile systems studied in this paper, $\zeta > 0$.

Equation (3) describes the evolution of the concentration of bacteria, where $\Gamma_\phi$ shows how fast $\phi$ responds to gradients in the chemical potential $\mu$. The first term in the free energy density (5) stabilises a nematic phase inside the drop (where $\phi \neq 0$). The second and third terms stabilise the circular droplet and the fourth term is the energy cost due to distortions in the nematic field, assuming a single Frank elastic constant $K_Q$. Finally, the last term is an anchoring free energy which favours parallel anchoring of the director at the interface for $a < 0$.

In the experiments the nematic alignment of the bacteria decreases towards the centre of the drop. To mirror this in the simulations the magnitude of the nematic order inside the drop is set by a field $\mathcal{J}$ with dynamics

$$\partial_t \mathcal{J} = D_s \nabla^2 \mathcal{J} - k_s \phi, \tag{8}$$

and with the boundary condition that $\mathcal{J} = 1$ outside the drop and also at the interface if $\phi < 0.5$. This dynamics results in the magnitude of the order, and as a result the magnitude of the active stress, decaying towards the center of the drop. Inside the drop, the magnitude of the nematic order is related to the field $\mathcal{J}$ through the relation $S^2 = \mathcal{J}$, and outside the drop $S$ is equal to zero.

The equations of active nematohydrodynamics are solved using a hybrid lattice Boltzmann and finite difference method in a square box with periodic boundary conditions [1, 5].

## B. Simulation parameters

For the simulations of the circular drop in Fig. 4(a)-(c), Fig. 4 (f)-(g), and Video S6 in Supplementary Materials we used the parameter values: $D_s = 0.1$, $\rho = 40$, $\gamma = 0.3$, $k_s = 0.0002$, $\zeta = 0.01$, $\Gamma_f = 0.3$, $K_Q = 0.005$, $K_\phi = 0.01$, $C = 0.001$, $A = 0.1$, $\Gamma_\phi = 0.1$, $\lambda = 0$, $a = -0.005$. The size of the simulation box and the total time of the simulation were $L = 450$ and $t_{tot} = 5 \times 10^5$ in lattice Boltzmann units, and the radius of the drop was $r_0 = 150$. The simulations were initialised with the director parallel to the interface and with a uniform noise in the interval $[0, 3\pi] \times 10^{-2}$. To calculate the director and the average velocity field in Fig. 4(b) we fixed the edge of the drop once the protrusions and the defects had formed and averaged the flow in the vicinity of sufficiently wide protrusions.

For the ellipse in Fig. 4(d)-(e) we chose the long and short axis equal to $r_2 = 150$ and $r_1 = 85$. We added an initial noise uniformly distributed in the interval $[0, \pi] \times 10^{-3}$. All the other parameters were the same as above. In Fig. 4(e), we averaged the velocity field at the scale of each protrusion when small protrusions form. The curvature is the curvature of the original ellipse.

---